\documentclass[10pt,onecolumn]{article}
\usepackage[mathscr]{euscript}
\usepackage{amssymb}
\usepackage{amsmath}
\usepackage{graphics}
\usepackage{latexsym}
\usepackage{array}
\usepackage[activeacute,english]{babel}
\usepackage[latin1]{inputenc}

\usepackage[english]{babel}
\usepackage{graphicx,bm}
\usepackage{amssymb,color,amsmath}
\usepackage[colorlinks=true]{hyperref}

\newcommand{\ket}[1]{\ensuremath{\left|#1\right\rangle}}
\newcommand{\bra}[1]{\ensuremath{\left\langle#1\right|}}

\newcommand{\be}{\begin{equation}}
\newcommand{\ee}{\end{equation}}

\newcommand{\bn}{\begin{eqnarray}}
\newcommand{\en}{\end{eqnarray}}
\newcommand{\bq}{\begin{equation}}
\newcommand{\eq}{\end{equation}}

\newcommand{\bc}{\begin{center}}
\newcommand{\ec}{\end{center}}

\begin{document}

\title{Energy operator for non-relativistic and relativistic quantum mechanics revisited}
\author{J. A. S\'anchez-Monroy\footnote{Grupo de Campos y Part\'\i culas, Universidad Nacional de Colombia, Sede Bogot\'a, Colombia}, John Morales\footnote{Grupo de Campos y Part\'\i culas, Universidad Nacional de Colombia, Sede Bogot\'a, Colombia,
Centro Internacional de F\'\i sica Bogot\'a, Colombia} and Eduardo Zambrano\footnote{Centro Brasileiro de Pesquisas Fisicas - CBPF, 22290-180, Rio de Janeiro, R.J., Brazil}}
\maketitle
\begin{abstract}
For quantum mechanics of a charged particle in a classical external electromagnetic field
the momentum and Hamiltonian operators are gauge dependent. For overcome this difficulty
we reexamined the effect of a gauge transformation on Schr\"odinger and Dirac equations.
We show that the gauge invariance of the operator $H-i\hbar \frac{\partial}{\partial t}$
provides a way to find the energy operator from first principles. In particular, when the system has
stationary states the energy operator can be identified without ambiguities for non-relativistic
and relativistic quantum mechanics. Finally, we examine other approaches finding that in
the case in which the electromagnetic field is time independent, the energy operator obtained
here is the same as one recently proposed by Chen \textit{et al.} \cite{Chen2}.
\end{abstract}

\section{Introduction}

The physical description of nature must not depend on
arbitrariness arisen from mathematical formalism. Typically some non-physical
quantities (NPQ) are introduced in order to find a solution for the
equations which governs the dynamics of a given system.
In electrodynamics a gauge transformation of the electromagnetic
fields is a well known example of  such an arbitrariness. Classical
mechanics with electromagnetic fields, via Lorentz force, is gauge invariant
\cite{Ja9,ed}. However, there are apparent problems concerning the momentum
and Hamiltonian operators of the charged particle in quantum mechanics,
due to the fact that the expectation values of these two operators
are gauge dependent \cite{Chen2,Chen5}-\cite{st4}.
\par
Experimental results are gauge independent, therefore it is necessary to find a
gauge invariant quantity to represent physical energy.
In the literature, there are some suggestions
for this purpose. One of them consists in a gauge invariant \emph{energy operator}
which was proposed by Yang \cite{Yang1}. This approach frequently has been used in
literature \cite{KoY0}-\cite{Bec}. However, as we will show, it is possible to find
counterexamples in which this approach does not represent the energy.
A different proposal suggests to restricting the gauge function to be a sum of a purely
spatial function plus a linear function of time \cite{st1}-\cite{st4}. We will show that
this approach yield to contradictory results.
\par
On the other hand, finding a definition of gluon spin and orbital angular momentum has been a long standing problem
\cite{jaffe,ji}. Recently Chen \textit{et al.} proposed a gauge invariant decomposition
of the total nucleon angular momentum into quark and gluon constituents \cite{Chen3,Chen4}.
The decomposition is based on the separation of the electromagnetic potential into pure gauge and gauge
invariant parts, which simultaneously satisfies the requirement of gauge invariance and the relevant
commutation relations. With this decomposition a new energy operator was proposed in the non-relativistic
and relativistic quantum mechanics \cite{Chen2,Chen5,Chen6}.
\par
The purpose of this paper is of set forth the gauge invariance of quantum mechanics. For this we will
show, in section \ref{Sec2}, that the operator $H-i\hbar \frac{\partial}{\partial t}$ is gauge invariant,
even for relativistic quantum mechanics. At the same time we will show how to derive the gauge invariant
energy operator from first principles explicitly when the system has stationary states in section.
In section \ref{Sec3}, we will discuss other approaches. Additionally, we will demonstrate that for time
independent electromagnetic fields, our approach is consistent with the decomposition of the potential
into a pure gauge and a gauge invariant parts, proposed in \cite{Chen2,Chen5,Chen6},
is consistent with our approach. Finally we summarize our conclusions in section  \ref{Sec4}.


\section{Gauge invariance}\label{Sec2}

Let us describe the electrodynamics using the potentials $\bm{A}$ and $A^0$, the classical electrodynamics is invariant by the so-called gauge transformations
\begin{equation}  \label{tg}
\bm{A}\to\bm{A}^{\prime }=\bm{A}+\nabla\chi,\quad\quad A^0\to {A^0}^{\prime
}=A^0-\frac{\partial\chi}{\partial t},
\end{equation}
where the scalar function $\chi$ is the \emph{gauge function}.
Due to the fact that classical equations of motion are given in terms
of the fields, the transformation Eq. \eqref{tg} does not change
anything in the dynamics, i.e. classical theory is gauge invariant.
\par
The canonical momentum and Hamiltonian are defined by
\begin{equation}\label{ExpectVa}
\mathbf{p}=m\mathbf{v}+q\mathbf{A}, \ \ \ \text{and}
\ \ \ H=\frac1{2m}\left(\mathbf{p}-\frac{q}{c}\bm{A}\right)^2+qA^0,
\end{equation}
respectively. Here $q$ denotes the charge of the particle. The Hamiltonian is quantized if $\mathbf{p}$  is replaced by $-i\hbar\mathbf{\nabla}$.
After a gauge transformation Eq. \eqref{tg}, the expectation value of the operators in Eq. \eqref{ExpectVa} are transformed as:
\begin{eqnarray}\label{momenexp}
\bra{\Psi^{\prime }}\mathbf{p}^{\prime }\ket{\Psi^{\prime }}&=&\bra{\Psi}\mathbf{p}\ket{\Psi}+q\bra{\Psi}\nabla\chi\ket{\Psi} \\
\bra{\Psi^{\prime }}H^{\prime }\ket{\Psi^{\prime }}&=&\bra{\Psi}H\ket{\Psi}+q\bra{\Psi}\frac{\partial\chi}{\partial t}\ket{\Psi}
\end{eqnarray}
where $\ket{\Psi^{\prime }}=e^{iq\chi(\mathbf{r},t)/\hbar}\ket{\Psi}$. Notice that the expectation values of these two operators are gauge dependent. In order to remove the gauge dependence of the expectation value of canonical momentum, one introduces the gauge invariant operator,
\begin{equation}
\mathbf{P}=\mathbf{p}-q\mathbf{A}.
\end{equation}
It is straightforward to check that the expectation value of this operator is gauge invariant, since the expectation value of $q\mathbf{A}$
cancels the additional term in the right side of in Eq. (\ref{momenexp}). However, the commutators between the components of $\mathbf{P}$ are
\begin{equation}
[P^i,P^j]=-iq(\partial^i A^j-\partial^j A^i)=-iqF^{ij}.
\end{equation}
therefore $\mathbf{P}$ does not satisfy the Lie algebra of canonical momentum (i.e. $[P^i,P^j]=0$), then it cannot be the proper momentum operator \cite{Chen2,Chen6}.
\par
On the other hand, the dynamics of a quantum particle is described by its (complex) wavefunction $\Psi(\mathbf{r},t)$, and its evolution is determined by the equation \cite{GreinerRQM}:
\begin{equation}  \label{Schr}
H\Psi(\mathbf{r},t)=i\hbar\frac{\partial}{\partial t}\Psi(\mathbf{r},t),
\end{equation}
where $\hbar$ is the Planck constant and $H$ is the Hamiltonian. Denoting by $A=(A^0,\bm{A})$ to the external electromagnetic potential, for non-relativistic quantum mechanics we have the Hamiltonian
\begin{equation}\label{SchHam}
H_{S}=\frac1{2m}\left(\mathbf{p}-\frac{q}{c}\bm{A}\right)^2+qA^0,
\end{equation}
and so we obtain the known Schr\"odinger equation. Meanwhile, for relativistic quantum mechanics we have the Dirac
Hamiltonian
\begin{equation}\label{DicHam}
H_{D}=\vec{\alpha} \cdot \left(\mathbf{p}-\frac{q}{c}\bm{A}\right)+\beta m+qA^0,
\end{equation}
where $\vec{\alpha}$ and $\beta$ are $4\times4$ Dirac matrices.
\par
Let us consider the effect of a unitary transformation $U(\chi)$ on Eq. \eqref{Schr},
\begin{equation}  \label{SchrPsi}
i\hbar U(\chi)\frac{\partial}{\partial t}{\Psi}=U(\chi)H\Psi.
\end{equation}
From the Leibniz's law for derivatives we have that	
\begin{equation}
U(\chi)\frac{\partial\Psi}{\partial t}=\frac{\partial}{\partial t}\left[%
U(\chi)\Psi\right]-\frac{\partial U(\chi)}{\partial t}\Psi.
\end{equation}
Thus Eq. \eqref{SchrPsi} which becomes
\begin{multline}  \label{SchPri}
i\hbar\frac{\partial}{\partial t}\left[U(\chi)\Psi\right]=U(\chi)H\Psi+i\hbar%
\frac{\partial U(\chi)}{\partial t}\Psi=\left[U(\chi)HU^\dag(\chi)+i\hbar\frac{\partial U(\chi)}{\partial t}%
U^\dag(\chi)\right](U(\chi)\Psi).
\end{multline}
Let us define the transformed wavefunction as $\Psi^{\prime }(\mathbf{r}%
,t)\equiv U(\chi)\Psi(\mathbf{r},t)$, hence the transformed Hamiltonian is given
by
\begin{equation}  \label{newha}
H^{\prime}\equiv U(\chi)HU^\dag(\chi)+i\hbar\frac{\partial U(\chi)}{\partial t}U^\dag(\chi)
\end{equation}
in order to obtain the transformed Schr\"odinger (or Dirac) equation
\begin{equation}  \label{SchrPsi2}
H^{\prime }\Psi^{\prime }=i\hbar\frac{\partial \Psi^{\prime }}{\partial t},
\end{equation}
Therefore, a unitary transformation $U(\chi)$ does not affect the
form of the Schr\"odinger (or Dirac) equation and the notion of probability,
since $|\Psi^{\prime }(\mathbf{r},t)|^2=|\Psi(\mathbf{r},t)|^2$. However, the
Hamiltonian has been changed by Eq. \eqref{newha}, which means that it
is not measurable and cannot be seen as an energy operator \cite{Golm}.
The particular choice of $U(\chi)=e^{iq\chi(\mathbf{r},t)/\hbar}$
reproduces the gauge transformations Eq. (\ref{tg}). So the Hamiltonian is modified
by the gauge transformation according to Eq. (\ref{newha}) and it does not constraint that their
eigenvalues remain unchanged. Consequently, the eigenvalue problem of the Hamiltonian, in general,
does not give the energy spectrum: The set of eigenvalues of $H$, denoted by $\alpha_n$,
are different from the energy levels of the system. This set is defined as by the equation
\begin{equation}  \label{valorH}
H(A^0,\bm{A})u_n=\alpha_nu_n,
\end{equation}
where $u_n$ are the eigenfunctions of $H(A^0,\bm{A})$. Notice that $\{\alpha_n\}$ in general are not the energy.
In order to clarify the ideas above enunciated,
consider the equation of eigenvalues of $H^{\prime }$,
\begin{equation}  \label{valorH'}
H^{\prime }({A^0}^{\prime },\bm{A}^{\prime })v_j({\mathbf{x}},t)=\beta_jv_j({%
\mathbf{x}},t).
\end{equation}
The sets $\{\alpha_k\}_{k\in I}$ y $\{\beta_j\}_{j\in I^{\prime
}}$, are not necessary the same and so the index sets $I$ and
$I^{\prime }$ may be different (e.g., a spectrum may be discrete
and the other not). Then, omitting the functional dependencies,
\begin{equation}
H^{\prime }({A^0}^{\prime },\bm{A}^{\prime })v_j= \left[UH(A^0,\bm{A}%
)U^\dag+i\hbar\frac{\partial U}{\partial t}U^\dag\right]v_j=\beta_j v_j,
\end{equation}
and so
\begin{equation}
\left[H+i\hbar U^\dag(\chi)\frac{\partial U(\chi)}{\partial t}\right]%
(U^\dag(\chi)v_j) =\beta_j(U^\dag(\chi)v_j).
\end{equation}
Now, if we define $f_j({\mathbf{x}},t)\equiv U^\dag(\chi) v_j({\mathbf{x}},t)$ and remarking that $U=e^{iq\chi(\mathbf{r},t)/\hbar}$, one has
\begin{equation}  \label{Hamiltopseudo}
\left[H-q\frac{\partial \chi({\mathbf{x}},t)}{\partial t}\right]%
f_j=\beta_jf_j.
\end{equation}
Note that the eigenvalues $\beta$ in Eqs. \eqref{valorH'} and \eqref{Hamiltopseudo} are the same, thus
$H^{\prime }$ and $H-q\frac{\partial \chi({\mathbf{x}}%
,t)}{\partial t}$ are equivalent operators\footnote{A linear operator $C$ in a Hilbert space $\mathscr{H}$ and a linear operator $C'$  in a Hilbert space $\mathscr{H}'$ are called \emph{equivalent}
if there exist an isomorphism $U$ of $\mathscr{H}$ onto $\mathscr{H}'$ (or an  automorphism $U$ of $\mathscr{H}$ if $\mathscr{H}=\mathscr{H}'$) such that
$C=U^{-1}C'U$ \cite{Gil1}.}. However, $H$ and $%
H^{\prime }$ are not equivalent operators, thus they do not have the same
eigenvalues. We conclude that in general the Hamiltonian does not represent
the energy. On the other hand, the procedure performed to
obtain Eq. \eqref{SchrPsi2} from Eq. \eqref{SchrPsi} ensures that the operator $%
H-i\hbar \frac{\partial}{\partial t}$ is equivalent to $H^{\prime }-i\hbar
\frac{\partial}{\partial t}$. This equivalence implies that the solutions
of the Schr\"odinger (or Dirac) equation for $H$ and $%
H^{\prime }$ are the same. In this sense, the quantum mechanics of a charged particle
in a classical external electromagnetic field is gauge invariant.
\par
So far we had demonstrated the gauge invariance of quantum mechanics. Now some questions rest: If the Hamiltonian does not represent the energy, which operator does? Can a time-dependent gauge transformation change the energy spectrum?
\par
First, let us review the recipe to obtain the energy spectrum of a particular system.
Suppose that Eq. \eqref{Schr} allows the separation of variables as
\begin{equation}  \label{SepaVar}
\Psi_k(\mathbf{r},t)=e^{-i\alpha_kt/\hbar}\psi_k(\mathbf{r}),
\end{equation}
thus Schr\"odinger (or Dirac) equation is read as the eigenvalues equation of $H(A^0,\bm{A})$
\begin{equation}  \label{Hpsi_k}
H(A^0,\bm{A})\psi_k(\mathbf{r})=\alpha_k\psi_k(\mathbf{r}).
\end{equation}
These $\psi_k(\mathbf{r})$ correspond to the stationary states of $H$. After
applying a transformation Eq. \eqref{tg}, the transformed wavefunction is
\begin{equation}
\Psi_{k\chi}(\mathbf{r},t)=e^{\frac{iq}\hbar\chi(\mathbf{r}%
,t)}\Psi_k(\mathbf{r},t)=e^{\frac{i}{\hbar}\left(q\chi(\mathbf{r},t) - \alpha_k t\right)}\psi_k(\mathbf{r})
\end{equation}
and the transformed Hamiltonian is $H^{\prime }(\bm{A}^{\prime },{A^0}^{\prime
})=H(A^0-\partial_t\chi,\bm{A}+\nabla\chi)$.
Notice that the transformed wavefunction is also stationary, because $|\Psi_{k\chi}(\mathbf{r}%
,t)|=|\psi_{k}(\mathbf{r})|$. If we substitute these transformed wavefunction and Hamiltonian in the
Schr\"odinger (or Dirac) equation Eq. (\ref{Schr}), it is easy to check that the time-independent eigenvalue equation associated is
\begin{equation}  \label{casi}
\left[H\left(A^0-\frac{\partial \chi}{\partial t},\bm{A}\right)+q\frac{%
\partial \chi}{\partial t}\right]\psi_k(\mathbf{r})=\alpha_k\psi_k(\mathbf{r}%
).
\end{equation}
We see that the left-hand side of the above equation is equal to left-hand side of the Eq. (\ref{Hpsi_k})
\begin{equation}
H\left(A^0-\frac{\partial \chi}{\partial t},\bm{A}\right)+q\frac{%
\partial \chi}{\partial t}=H(A^0,\bm{A}),
\end{equation}
and therefore Eqs. \eqref{casi} and \eqref{Hpsi_k} have the same solutions.
Therefore, the operator obtained from the separation
of variables is gauge invariant and represents the energy.
We conclude that in the particular case in which the Schr\"odinger (or Dirac) equation
allows a separation of variables as in Eq. (\ref{SepaVar}), the energies are given by the eigenvalues
$\alpha_k$ of the Hamiltonian whose eigenfunctions are a product of an spatial function by $e^{-i\alpha_k t}$.
Additionally, if another Hamiltonian ($H^{\prime }$) is related with the \emph{separable} one ($H$) by means
a gauge transformation, the eigenvalue equation of $H^{\prime }$ must be modified by introducing the term $\partial/\partial t$,
in order to perform the separation of variables. Hence this new \emph{eigenvalue equation} in terms of  $H^{\prime }$
(Eq. (\ref{casi})) provides the actual energies.
\section{Discussion about other interpretations}\label{Sec3}
As we mentioned in the Introduction, there are other attempts to resolve the gauge invariance of
quantum mechanics, though, as we will show in this section, some of these approaches lead to wrong conclusions,
whereas thee most recent proposal by Chen \emph{et al} \cite{Chen2,Chen5,Chen6,Chen3,Chen4} is consistent with
our present approach.
\subsection{Yang's energy-operator}
First we will consider the Yang's energy-operator, then we will discuss the restrictions on the gauge function
\cite{st1} and we will finish by explaining the approach of Refs. \cite{Chen2,Chen5,Chen6,Chen3,Chen4}.
\par
The fact that the Hamiltonian is not a gauge
invariant quantity, and so a NPQ, strikes against the common
meaning of the Hamiltonian as the energy. A different proposal
is given by Yang and it has been used in literature \cite{Yang1,KoY0,Ko2,Ber,KoY1,Nagu,Qi2,Klai,Qi1,Bec}:
in order define a gauge invariant operator representing the energy is introduced the \emph{Yang's energy-operator}
$\mathcal{Y}(\bm{A})$,
\begin{equation}  \label{enerope}
\mathcal{Y}(\bm{A})=H(A^0,\bm{A})-qA^0.
\end{equation}
A simple application of Eq. \eqref{newha} shows its invariance.
However, a typical system, as the Hydrogen atom denies its plausibility.
For instance, consider the $n$-th level of an Hydrogen atom
\begin{equation}
E_n =-{\frac{1}{2}}\left( {\frac{m{e}^4}{(4\pi\epsilon_0\hbar)^2}} \right ) {\frac{1}{n^2}}.
\end{equation}
This standard result is obtained by using the Hamiltonian of the form $%
H(A^0,\bm{A}=0)$, with an electrostatic potential $A^0(r)=-e/4\pi\epsilon_0|\mathbf{r}|$.
This Hamiltonian does not depend on time, and so the solutions of its
time-dependent Schrodinger equation may be calculated by using a separation
of variables
\begin{equation}
\Psi(\mathbf{r},t)={\ e}^{-i{E_n}/{\hbar}t}\psi(\mathbf{r}).
\end{equation}
For this case, the eigenvalue equation of the Yang's energy-operator is:
\begin{equation}\label{ondapla}
\mathcal{Y}(\bm{A}=0)\psi(\mathbf{r},t)=[H(A^0,\bm{A}=0)-eA^0]\psi(\mathbf{r},t)
=\frac{\mathbf{p}^2}{2m}\psi(\mathbf{r})=\epsilon\psi(\mathbf{r},t).
\end{equation}
Being $\epsilon$ the eigenvalues of $\mathcal{Y}(\bm{A})$. The above expression is obviously a free-particle Hamiltonian, it
means that the spectrum is continuous!. Notice that Eq. (\ref{enerope}) is equal to the free-particle Hamiltonian for any $H=\mathbf{p}^2/2m+eA^0(x)$.
\par
Now, if we start by taking the temporal gauge for the Hydrogen atom, we have that
\begin{equation}
A^0=0, \ \ \text{and } \ \ \ \bm{A}=\frac{e t}{r^3}\mathbf{r},
\end{equation}
thus, the eigenvalue equation of $\mathcal{Y}$ is
\begin{equation}\label{enera}
\mathcal{Y}(\bm{A})\psi(\mathbf{r},t)=[H(A^0=0,\bm{A})]\psi(\mathbf{r},t)
=\frac{(\mathbf{p}-e\mathbf{A})^2}{2m}\psi(\mathbf{r})=\epsilon\psi(\mathbf{r},t).
\end{equation}
Now if we multiply both sides of Eq. (\ref{enera}) by $U(\chi)={e}^{ie\chi(\mathbf{r}%
,t)/\hbar}$, with $\chi(\mathbf{r},t)=et/r$ and taking into
account the identity
\begin{equation}\label{ident}
e^{ie\chi(\mathbf{r},t)/\hbar}\frac{({\mathbf{p}}-e{\bm{A}})^2}{2m}\psi(%
\mathbf{r},t)=\frac{\left[{\mathbf{p}}-e({\bm{A}}+\nabla\chi)\right]^2}{2m}e^{ie\chi(\mathbf{r}%
,t)/\hbar}\psi(\mathbf{r},t),
\end{equation}
the eigenvalues equation becomes
\begin{equation}
\frac{\mathbf{p}^2}{2m}\left[e^{ie\chi(\mathbf{r},t)/\hbar}\psi(%
\mathbf{r},t)\right]=\epsilon\left[e^{ie\chi(\mathbf{r},t)/\hbar}\psi(%
\mathbf{r},t)\right].
\end{equation}
Again we obtain a similar result as in Eq. (\ref{ondapla}). The above arguments
proof the unsuitability of the energy-operator to describe properly the energy.
Specifically, we proved that the operator $\mathcal{Y}(\bm{A})$ in the Eq. \eqref{enera}
is equivalent to operator $\frac{\mathbf{p}^2}{2m}$, and so they have the same eigenvalues. Consequently the
operator $\mathcal{Y}(\bm{A})$ in Eq. (\ref{enerope}) is not the energy in general.
\subsection{Restrict the gauge function}
\begin{figure}[h!]
\centering
\includegraphics[scale=0.3]{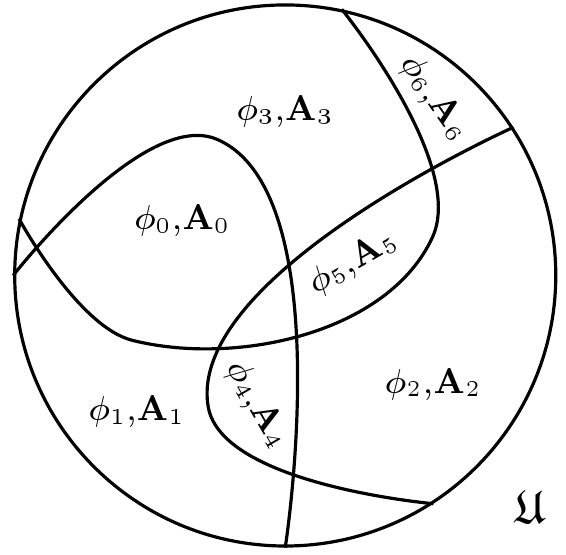}  \centering
\caption{Potentials space $\mathfrak{U}$. Each piece of this
circle corresponds to a set of potentials that are related between
themselves by Eq. \eqref{forma}. For each set we may chose a
pair ($A^0,\mathbf{A}$) to represent it, dividing $\mathfrak{U}$
on equivalence classes (sets of potentials). (We
here show few classes, but actually are infinite).} \label{haus1}
\end{figure}

Another approach, proposed by Stewart \cite{st1}, consists in restricting the gauge function to the form:
\begin{equation}  \label{forma}
\chi(\mathbf{r},t)=f(\mathbf{r})+g(t).
\end{equation}
This constraint is equivalent to assume the separation of the operator:
\begin{equation}
U(\chi)=e^{iqf(\mathbf{r})/\hbar}e^{iqg(t)/\hbar}.
\end{equation}
However, allowing this restriction leads to ambiguity when choosing the gauge potential.
To see this, let us define $\mathfrak{U}$ as a set
of all possible electrodynamic potentials that \emph{may represent}
a given system. Eq. \eqref{forma} divides the set $\mathfrak{U}$ in
equivalence classes (see fig. \ref{haus1}).
According to Stewart any two Hamiltonians, $H$ and $H'$, yields the same
\emph{physical results} only if they are related by a gauge transformation of the
form \eqref{forma}. Under this assumption, if we take two different equivalence classes,
we obtain two different physical results and as there is no way to choose beforehand
which is the ``real'' equivalence class that represents the physical system.
Then the restriction on the gauge function, Eq. (\ref{forma}), does not gives a satisfactory
solution for the gauge invariance, because it does not decide which equivalence class describes
properly the physical system.
\subsection{Decomposition of the electromagnetic potential}
Finally, let us consider a decomposition of the electromagnetic potential
into pure gauge and gauge invariant parts \cite{Chen2,Chen5,Chen6,Chen3,Chen4},
\begin{equation}\label{decom}
\mathbf{A}=\mathbf{A}_{pure}+\mathbf{A}_{phys}, \ \ \ \ \ A^0=A^0_{pure}+A^0_{phys},
\end{equation}
with
\begin{equation}\label{decom1}
\mathbf{\nabla}\times\mathbf{A}_{pure}=0, \ \ \ \ \ \mathbf{\nabla}\cdot\mathbf{A}_{phys}=0,
\end{equation}
\begin{equation}\label{decom2}
\nabla A^0_{pure}=-\frac{1}{c}\frac{\partial\mathbf{A}_{pure}}{\partial t}, \ \ \ \ \ \mathbf{\nabla}^2A^0_{phys}=-\frac{\rho}{\epsilon_0}.
\end{equation}
Under a gauge transformation Eq. (\ref{tg}) these two parts are transformed as follows,
\begin{equation}
\mathbf{A}_{pure}^{\prime}=\mathbf{A}_{pure}+\nabla\chi, \ \ \ \ \ \mathbf{A}_{phys}^{\prime}=\mathbf{A}_{phys}, \ \ \ \ \ A^{0\prime}_{pure}=A^0_{pure}-\frac{\partial\chi}{\partial t}, \ \ \ \ \  A^{0\prime}_{phys}=A^0_{phys}.
\end{equation}
In this approach the energy operator is $H(A^0,\mathbf{A})-qA^0_{pure}=H(A^0_{phys},\mathbf{A})$,
where $H=H_S$ for non-relativistic quantum mechanics, and $H=H_D$ for
relativistic quantum mechanics\footnote{It can be proved that the operator
$H(A^0_{phys},\mathbf{A})$ in \cite{Chen2} is the same energy operator
$H(A^0,\mathbf{A})-\frac{q}{c}\partial_t \frac{1}{\nabla ^2}\nabla \cdot \mathbf{A}$ in \cite{Chen3,Chen4},
using that $\nabla A^0_{pure}=-\frac{1}{c}\partial_t \mathbf{A}_{pure}$, one has $A^0-\frac{q}{c}\partial_t \frac{1}{\nabla^2}\nabla
\cdot \mathbf{A}=A^0_{phys}$.}. It is straightforward to prove that $H-eA^0_{pure}$ is gauge independent \cite{Yang1}.
\par
Next we will show that the energy operator $H(A^0_{phys},\mathbf{A})$ is consistent with our analysis of first principles, at least for the
particular case of a time independent electromagnetic field. In this case
\begin{equation}
\mathbf{E}(\mathbf{r})=-\nabla A^0-\frac{1}{c}\frac{\partial\mathbf{A}}{\partial t}, \ \ \ \ \ \mathbf{B}(\mathbf{r})=\nabla\times\mathbf{A}.
\end{equation}
Without loss of generality we can choose $\mathbf{A}(\mathbf{r})$ and $A^0(\mathbf{r})$  to be also time independent, because the magnetic field is time independent. By means of introducing the decomposition Eq. (\ref{decom}), and using Eq. (\ref{decom1}) one can derive
\begin{equation}
\mathbf{E}(\mathbf{r})=-\nabla A^0(\mathbf{r}), \ \ \ \ \ \mathbf{B}(\mathbf{r})=\nabla\times\mathbf{A}_{phys}(\mathbf{r}),
\end{equation}
on the other hand using Eq. (\ref{decom2}), $\nabla A^0_{pure}=0$ and implies that $A^0_{pure}=$ const. This constant can always be absorbed by $A^0_{phys}$. Then, Eq. (\ref{Schr}) can be written as
\begin{equation}
H(A^0_{phys}(\mathbf{r}),\mathbf{A}(\mathbf{r}))\Psi(\mathbf{r},t)=i\hbar\frac{\partial}{\partial t}\Psi(\mathbf{r},t),
\end{equation}
that allows as to have variable separation of the kind of Eq. (\ref{SepaVar}). The energy operator obtained their
$H(A^0_{phys}(\mathbf{r}),\mathbf{A}(\mathbf{r}))$, is the same as one recently proposed by Chen \textit{et al.} \cite{Chen2,Chen5,Chen6}.


\section{Conclusions}\label{Sec4}
In this paper we have reexamined the gauge invariant of non-relativistic and relativistic
quantum mechanics of a charged particle in a classical external electromagnetic field.
We show how the gauge invariance of the operator $H-i\hbar \frac{\partial}{\partial t}$
provides a way to find the energy operator. In particular, when the system has stationary states
the energy operator can be identified without ambiguities from first principles. We have dismissed some approaches
suggested in the literature and we have found that the energy operator obtain in this paper
(for time independent electromagnetic field) is the same one recently proposed by
Chen \textit{et al.} \cite{Chen2,Chen5,Chen6}.

\section*{Acknowledgements}
The authors would like to thank Y. Pérez for their useful comments. One of us (E.Z.)
thanks partial financial support from Faperj (Brazilian agency).

\end{document}